\documentclass[12pt]{iopart}
\usepackage[utf8]{inputenc}
\usepackage{graphicx}
\usepackage{multirow}
\usepackage{color}
\usepackage{mcite}

\begin{document}

\title[]{Incommensurate magnetic ordering in CrB$_2$}
\author{A Deák$^1$, J Jackson$^2$, B Nyári$^{1,3}$, L Szunyogh$^{1,3}$}

\address{$^1$ Department of Theoretical Physics, Institute of Physics, Budapest University of Technology and Economics, M\H uegyetem rkp.\ 1-3., H-1111 Budapest, Hungary}
\address{$^2$ STFC Scientific Computing Department, Daresbury Laboratory, Warrington WA4~4AD, UK}
\address{$^3$ ELKH-BME Condensed Matter Research Group, Budapest University of Technology and Economics, M\H uegyetem rkp.\ 1-3., H-1111 Budapest, Hungary}

\ead{nyari.bendeguz@ttk.bme.hu}

\begin{abstract}
Incommensurate magnetism in CrB$_2$ is studied in terms of a spin model based on density functional theory calculations. Heisenberg exchange interactions derived from the paramagnetic phase using the disordered local moment theory show significant differences compared with those resulting from the treatment of the material as a ferromagnet; of these two methods, the disordered local moment theory is found to give a significantly more realistic description.
We calculate strongly ferromagnetic interactions between Cr planes but largely frustrated interactions within Cr planes.
Although we find that the ground state ordering vector is sensitive to exchange interactions over a large number of neighbour shells, the $q$-vector of the incommensurate spin spiral state is satisfactorily reproduced by the theory ($0.213$ compared with the known ordering vector $0.285\times (2\pi)/(a/2)$ along $\Gamma-{\rm K}$).  The strong geometric frustration of the exchange interactions
 causes a rather low Néel temperature (about 97 K), also in good agreement with experiment.
\end{abstract}

\vspace{2pc} \noindent{\it Keywords}: incommensurate spin spiral, Heisenberg model, disordered local moment, ab initio, frustration

%
%
%
%

\section{Introduction}

Beside geometric frustration stemming from an incompatibility of the underlying lattice with magnetic interactions, frustration caused by competing exchange couplings is also known to lead to non-collinear magnetic ordering. Not only can it cause spin spiral ground states in atomic chains \cite{laszloffy2019,laszloffy2021} and surfaces \cite{rozsa2015}, it can also contribute to the stabilization of skyrmions \cite{dupe2014,dupe2016} and induce an effective attractive interaction between them \cite{rozsa2016}. Frustrated interactions may even lead to incommensurate magnetic ordering in bulk systems, for which a typical example is the long-wavelength helical order of Mn$_3$Sn \cite{park2018}.

\begin{figure}
  \begin{centering}
    \includegraphics[width=\linewidth]{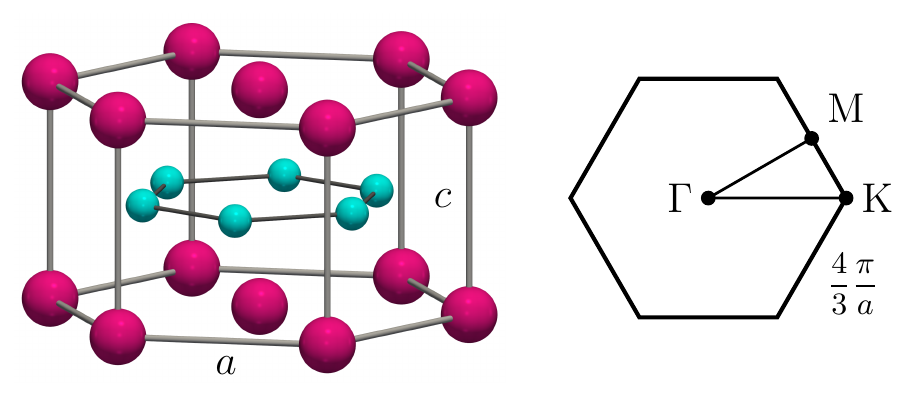}
  \end{centering}
  \caption{\label{fig:crb2-crystal}Crystal structure of CrB$_2$. (Left) The direct lattice, where large pink and small cyan spheres denote Cr and B atoms, respectively. The lattice parameters $a$ and $c$ are also indicated. (Right) The hexagonal two-dimensional Brillouin zone for $q^z=0$. Corresponding to ${\rm C}_{3v}$ symmetry the irreducible wedge is bounded by the $\Gamma - {\rm K} - {\rm M}$ triangle.}
\end{figure}

Among transition metal diborides CrB$_2$ is another challenging example of an itinerant antiferromagnetic (AFM) metal possessing incommensurate magnetic ordering.
It crystallizes in the hexagonal AlB$_2$ (or $C32$) crystal structure with space group P6/mmm \cite{portnoi1969}, in which honeycomb layers of boron alternate with triangular chromium layers, see \fref{fig:crb2-crystal}.
The magnetic structure was determined experimentally by neutron scattering by Funahashi and coworkers \cite{funahashi1977}, revealing that the Cr moments order antiferromagnetically with an incommensurate spin spiral in the Cr planes with ordering vector $0.285 \times (2\pi)/(a/2)$ along the $\langle 110 \rangle$ direction, i.e.\ a fraction of 0.86 along the distance from $\Gamma$ to $\rm K$ (the direction $\Lambda$),
and with Néel temperature 88 K; this description has been extended by recent experiments~\cite{park2020}, where the spin-wave spectrum was measured across a large $\left(\mathbf q, \omega\right)$-window.
A detailed investigation into the nature of magnetic ordering in CrB$_2$ was provided by Kaya \emph{et al.}~\cite{kaya2009}, which confirmed that Cr moments form a simple helix structure, rotating in the plane formed by the $c$-axis and $\langle 110\rangle$ (the propagation direction).
An examination of the temperature-dependent conductivity of CrB$_2$ around the Néel temperature by Bauer \emph{et al.}~\cite{bauer2014} included the determination of the susceptibility, showing Curie--Weiss susceptibility with a large, negative $\Theta_{\rm CW}$
indicative of strong frustration, together with a Cr moment of $\sim 2$ $\mu_{\rm B}$, in contrast to early neutron scattering experiments which indicated a Cr moment of $\sim 0.5$ $\mu_{\rm B}$~\cite{funahashi1977}.

The electronic structure of CrB$_2$ has been examined theoretically in a number of works based on band theory using the local density approximation or generalised gradient approximation (GGA) for the exchange-correlation in density functional theory.
Early calculations using non-self-consistent potentials~\cite{liu1975} have been supplemented by linear muffin-tin orbital (LMTO)~\cite{vajeeston2001,grechnev2009} and later full-potential augmented plane-wave calculations.
Of these more accurate GGA calculations, application of the generalised Bloch theorem (i.e.\ ``spin-spiral'' calculations) showed ordering along $\mathbf{q} \approx 0.3 \, \mathbf{q}_{110}$ and Cr moment $1.3$ $\mu_{\rm B}$ as part of detailed description of the de Haas--van Alphen effect~\cite{brasse2013}.

The present work is aimed at extending our understanding of the magnetism in CrB$_2$ in terms of the Heisenberg exchange interactions leading to incommensurate ordering, both in the low-temperature limit, which we model as a ferromagnet (FM), and in the high-temperature limit where the local Cr moments are treated as fully disordered. First, we give brief details of the computational methods we used: the electronic structure calculations are based on Green's function (GF) multiple scattering formalism which facilitates the evaluation of magnetic exchange interactions and the study of the paramagnetic phase using the disordered local moment (DLM) method \cite{gyorffy1985}. Then we present our results for the exchange interactions in both the FM and DLM states and give mean-field estimates for the wave vector ${\bf q}_0$ of the spin spiral ground state and the Néel temperature $T_{\rm N}$. These values are recalculated by spin dynamics and Monte Carlo simulations, respectively. We conclude that the DLM based spin models give more realistic description of the magnetism of CrB$_2$ than the spin model derived from the ordered (FM) state.

\section{Details of calculations}
Calculations have been performed using
the Screened Korringa--Kohn--Rostoker (SKKR) method~\cite{kkrbook}
in the atomic-sphere approximation (ASA).
The ASA describes the crystal potential by a sum of overlapping atomic spheres, neglecting the interstitial region;
this approximation is appropriate for close-packed structures like AlB$_2$ materials including CrB$_2$.
The ASA constraint, that the sum of sphere volumes equals the unit cell volume, does not determine the relative size of Cr and B spheres.  We have experimented with sphere sizes obtained by minimizing the average sphere overlap and by scaling up touching spheres defined by the first maximum of the electrostatic potential and found that both configurations give closely similar results.
For all calculations we assumed an in-plane lattice parameter of $a=2.972$ \AA\ and $c=3.066$ \AA\ according to experiments \cite{portnoi1969}.
The resulting ASA radii are therefore 
$S_{\rm Cr}= 1.580$ \AA\  and $S_{\rm B} = 0.929$ \AA .

All calculations used an angular momentum cut-off of $\ell_{\rm max}{=}2$ and the GGA exchange-correlation parameterisation as formulated by Perdew, Burke and Ernzerhof~\cite{perdew1996}.
16 energy points were used for complex contour integrations in energy, with 546 $k$-points in the two-dimensional Brillouin zone (2d BZ) for self-consistent-field calculations and up to 20\,000
near the Fermi energy for computing exchange interactions.
Ferromagnetic self-consistent calculations were combined with the Relativistic Torque Method (RTM) to derive exchange interactions using infinitesimal rotations of the spins~\cite{udvardi2003}.
For the paramagnetic phase, the relativistic development of the disordered local moment theory was used~\cite{gyorffy1985,staunton2006},
from which the adiabatic magnetic energy surface was mapped onto a spin model using the Spin-Cluster Expansion (SCE)~\cite{szunyogh2011}.

The classical spin model we extract with the above methods is of the form
\begin{equation}
  \mathcal H = 
  - \frac{1}{2} \sum\limits_{i\ne j} \mathbf S_i \underline{\underline J}_{ij} \mathbf S_j
  + \sum\limits_{i}\mathbf S_i \underline{\underline K}_{i}\mathbf S_i  \, , \label{eq:spin_model}
\end{equation}
where $i$ and $j$ both run over all Cr atoms and ${\mathbf S}_i$ stands for classical spins of unit magnitude.
Both the RTM and the SCE methods allow the extraction of the full $\underline{\underline J}_{ij}$ tensor, encompassing the isotropic exchange interaction, the Dzyaloshinskii--Moriya interaction and the two-site exchange anisotropy~\cite{udvardi2003}, the latter two ones originating from spin-orbit interaction. In the convention used in \eref{eq:spin_model} a positive isotropic interaction corresponds to FM coupling.
Note that the Dzyaloshinskii--Moriya interaction is absent due to inversion symmetry present between any two Cr atoms.
The magnetic anisotropy is uniaxial, favouring out-of-plane Cr ordering. However, since the spin-orbit interaction is very weak in this system, the exchange anisotropy and the on-site anisotropy related to the matrix $\underline{\underline K}$ are also very small, in the order of 10 $\mu$Ry, and thus aren't expected to significantly affect the ground state ordering.

We have repeated our calculations using the tight-binding LMTO Green's function (GF) method, as implemented in the Questaal code~\cite{pashov2020}.
Compared with the SKKR calculations, the LMTO calculations use the same ASA construction, but were scalar-relativistic and used a 3rd order parameterisation of the potential function, without inclusion of the combined correction for the potential overlap.
Self-consistent LMTO-GF calculations involved integration on an eliptical energy contour with 35 points and a BZ mesh of 49$\times$49$\times$41 points, by sampling.
We found that the results of the two methods are in very good agreement.

The spin model of \eref{eq:spin_model} has been studied using Monte Carlo and spin dynamics methods assuming classical statistics\cite{nowak2007}. We used Landau--Lifshitz--Gilbert spin dynamics simulations at zero temperature as a means of energy minimization for a ground-state search. For this we first verified that the ground-state spin configuration has $q^z=0$, then used a $64 \times 64$ Cr atom 2d lattice with free boundary conditions in-plane and periodic boundary conditions perpendicular to the plane. The simulations were converged in energy to below $10^{-8}$ Ry per Cr atom. For the temperature-dependence of the specific heat we used Monte Carlo simulations on a $32 \times 32 \times 32$ Cr atom lattice with free boundary conditions in order to estimate the Néel temperature. The last $3 \times 10^9$ time steps were considered for thermal averaging out of $5 \times 10^9$ total time steps.

\section{Results}
First we performed self-consistent calculations by using the SKKR method in the FM and DLM states.
For the FM state we obtained 1.564 $\mu_{\rm B}$ spin moment for the Cr and 0.045 $\mu_{\rm B}$ for the B atoms.
In the DLM self-consistent calculation the Cr spin moment decreased to 1.344 $\mu_{\rm B}$, and there was no moment
induced on the B sites due to the vanishing Weiss field in the paramagnetic state. This also justifies the softening of the Cr spin moment in the DLM state as compared to the FM state.  LMTO (scalar-relativistic) Cr moments are 1.577 and 1.301 $\mu_{\rm B}$ for the FM and DLM configurations, respectively.
The FM spin moment is significantly larger than in earlier spin-spiral calculations of Brasse \emph{et al.}~\cite{brasse2013} which can be explained by the larger volume of the atomic sphere defining the atomic moment than the non-overlapping spheres used in full-potential methods.

\begin{figure}
  \begin{centering}
    \includegraphics[width=\linewidth]{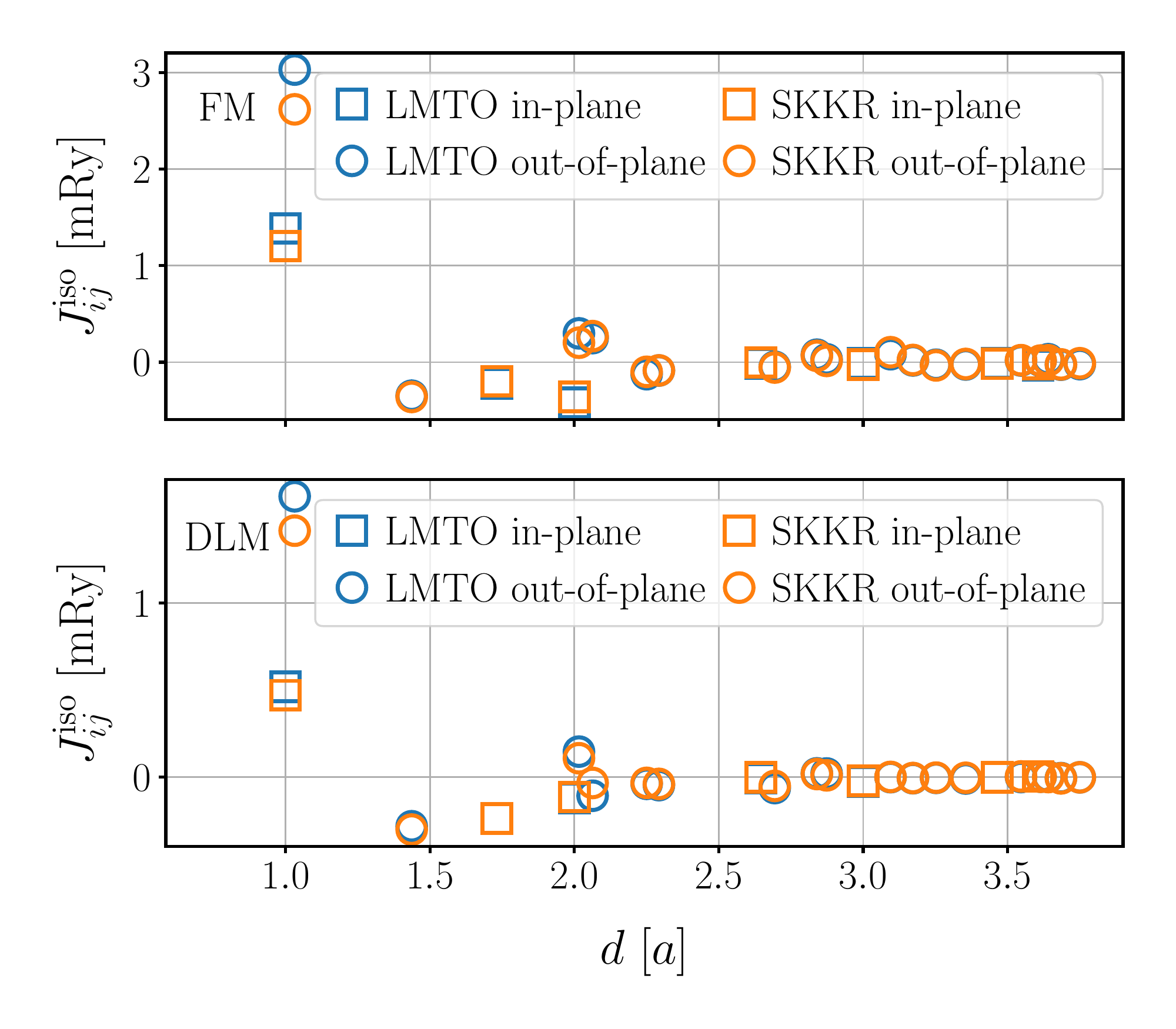}
  \end{centering}
  \caption{\label{fig:jiso-d}Isotropic Heisenberg couplings vs.\ interatomic distance. Parameters obtained with the SKKR (orange) and LMTO (blue) methods are compared for the FM (top) and DLM (bottom) reference states separately. Square and circle markers represent in-plane and out-of-plane neighbours, respectively.}
\end{figure}

We derived parameters for the spin model defined in \eref{eq:spin_model} in terms of the SKKR method by using the RTM for the FM configuration and using the SCE for the DLM state. Within the scalar relativistic LMTO method the calculation of only the isotropic exchange interactions was possible by using the method of infinitesimal rotations~\cite{LKAG1987,turek2006}.
A comparison of the resulting isotropic Heisenberg couplings is shown in \fref{fig:jiso-d}.
The two magnetic reference configurations produce broadly similar spatial distributions of spin model parameters, with the two nearest neighbour (NN) shells visibly dominating the interaction landscape.
In both cases the parameters obtained with SKKR and LMTO are in excellent agreement.
The main feature we note is that the exchange couplings obtained using the FM reference are about twice as large as those for the DLM,
which can partially be related to the larger magnitude of the Cr moment in the FM state.

With the definition in \eref{eq:spin_model} in mind,
it can be shown that the mean-field paramagnetic spin susceptibility can be related to the $\underline{\underline J}\left(\mathbf q\right)$ lattice Fourier transform of
the exchange tensors. In particular, the mean-field estimate predicts a magnetic ordering with highest critical temperature at the $\mathbf q_0$ wave vector for which
the maximal eigenvalue $J\left(\mathbf q\right)$ of the matrix $\underline{\underline J}\left(\mathbf q\right)$ is the highest within the BZ; the
corresponding ordering temperature is equal to $J\left(\mathbf q_0\right)/\left(3 k_{\rm B}\right)$. For a detailed derivation of this mean-field estimate see e.g.\ the
Appendix of Ref.~\cite{deak2011}.

The ordering wave vector from the mean-field estimates turned out in each case to lie along the $\Gamma - {\rm K}$ line connecting the centre of the hexagonal BZ with one of the
vertices with $q_0^z = 0$. We have found that despite the apparent dominance of the first two FM NN couplings there is enough frustration in the system to push magnetic
ordering away from the $\Gamma$ point. The competition of these strong FM couplings with AFM further neighbours results in a delicate balance ultimately giving rise to
the incommensurate spin spiral ground state possessed by CrB$_2$.

\begin{figure}[htb]
  \begin{centering}
    \includegraphics[width=\linewidth]{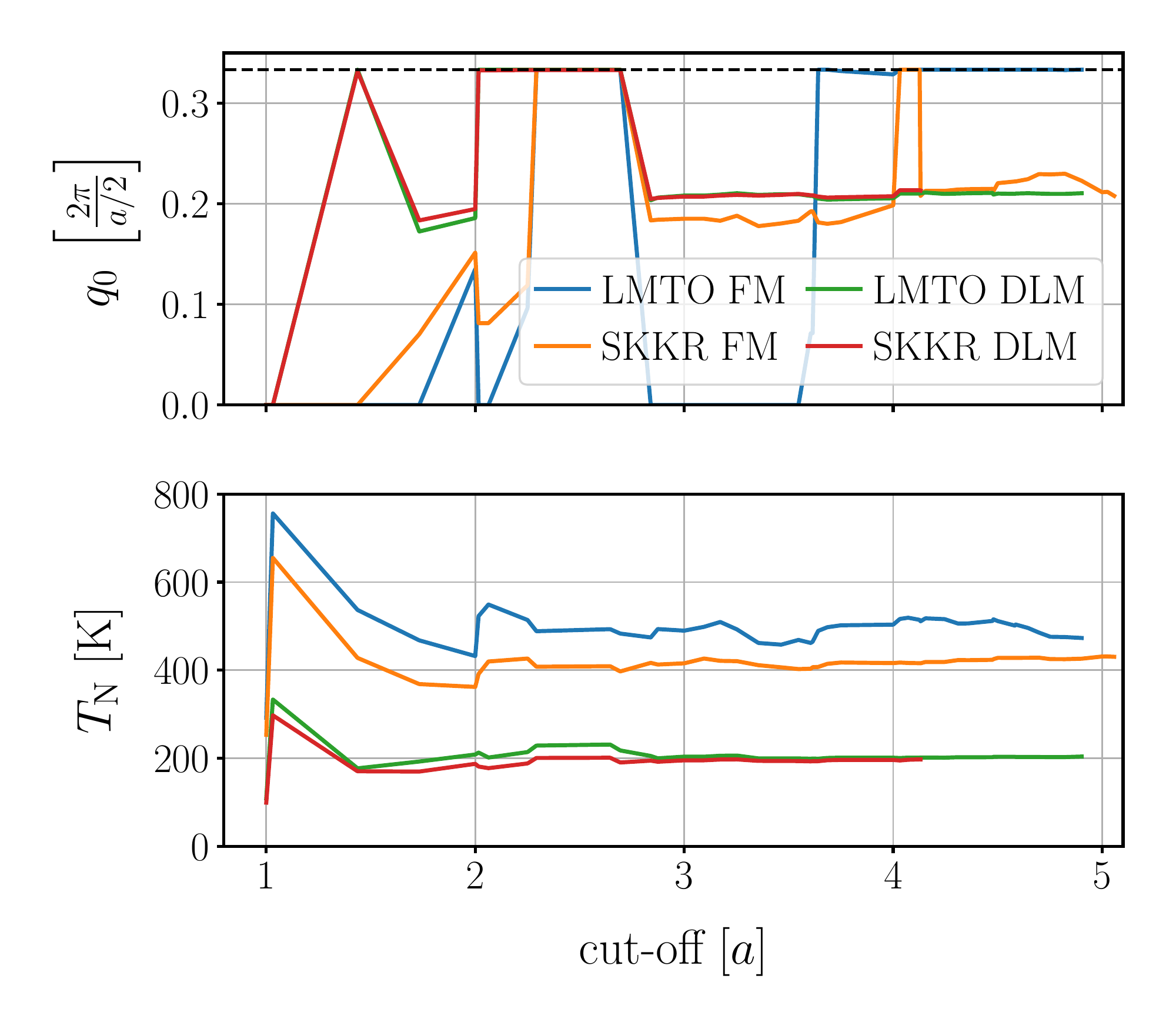}
  \end{centering}
  \caption{\label{fig:mean_field}(top) Magnitude of the mean-field ordering wave vector. $q=0$ corresponds to the $\Gamma$ point, whereas $q=\frac{1}{3} 2 \pi/(a/2)$, marked by the dashed line, corresponds to the ${\rm K}$ point of the BZ. (bottom) Mean-field Néel temperature as a function of interatomic cut-off.}
\end{figure}

In order to demonstrate this complex interplay of spin model parameters we computed the mean-field estimates while taking into consideration gradually more interacting neighbour shells.
\Fref{fig:mean_field} shows the dependence of the obtained mean-field ordering wave vectors as a function of this distance cut-off. Despite the magnitude of the
first two NN couplings the mean-field theory shows that pairs as far as 2.8 lattice constant units (12 shells) are necessary to achieve convergence in the DLM state.
This is even more so the case with the FM reference state, following the common observation that magnetic interactions tend to be shorter ranged in the DLM state due
to spin disorder. We computed exchange interactions in the FM state via SKKR for up to 9 lattice constant units to trace all distant interactions, and about $4.2 \, a$ distance
is necessary in order to achieve convergence. This extreme sensitivity to otherwise small exchange interactions is strong evidence for frustration.

We note that the sudden jumps in the mean-field $q_0$ estimate might first seem at odds with the fact that the corresponding distant interactions are small. However, these
jumps are merely the result of small changes along a nearly degenerate line of the $J\left(\mathbf q\right)$ function along the $\Gamma - {\rm K}$ line. When two local maxima
of the function have a crossing (with respect to their function values), the mean-field estimate for the ordering wave vector will abruptly jump from one maximum to the other.
It is clear in the bottom of \fref{fig:mean_field} presenting the mean-field Néel temperatures that there are no sudden changes beyond a cut-off of $2a$, where the last group of larger Heisenberg couplings are located.

\begin{figure*}[htb]
  \begin{centering}
    \includegraphics[width=\linewidth]{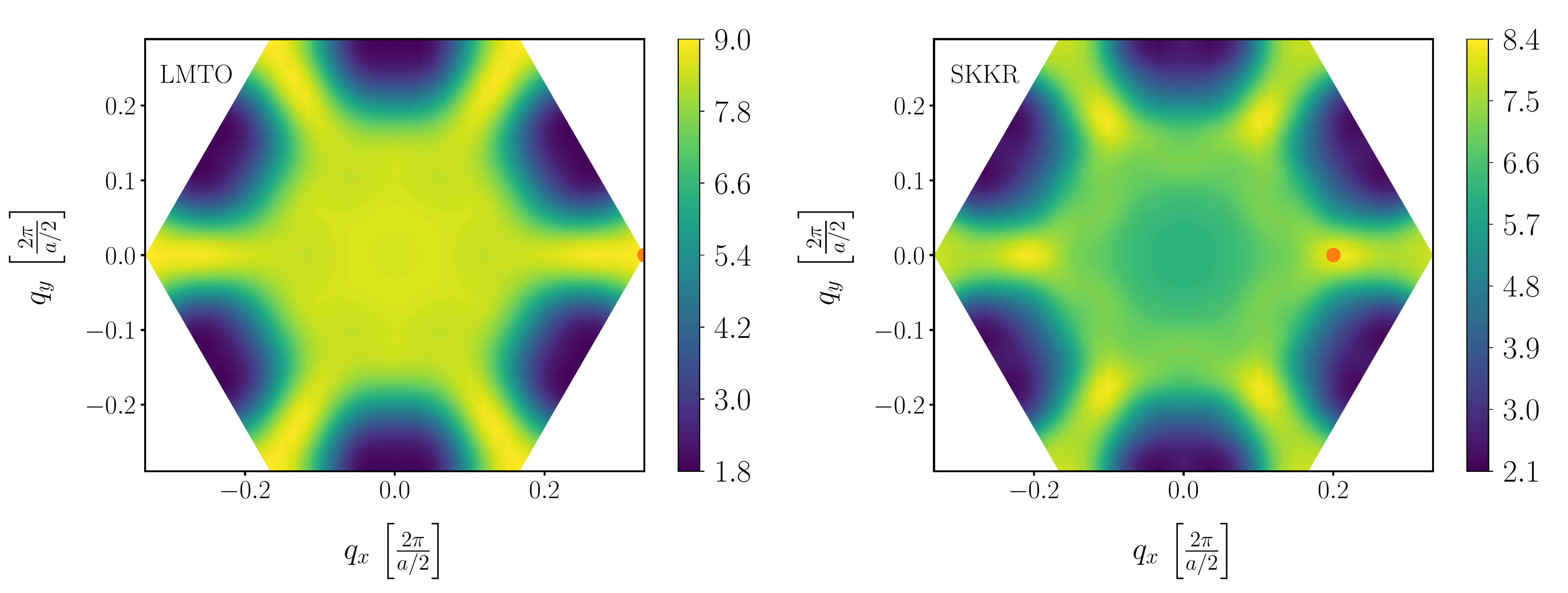}
  \end{centering}
  \caption{\label{fig:jq_surfaces}$J\left(\mathbf q\right)$ surface for $q^z=0$ in the 2d BZ for a FM reference state (left) using LMTO and (right) using SKKR-derived exchange interactions. The colorbars use mRy units. The vertices of the hexagonal 2d BZ (the ${\rm K}$ points) correspond to a triangular Néel state. The numerical maxima along a $\Gamma - {\rm K}$ line are indicated with orange circles.}
\end{figure*}

The specific mean-field estimates are collected in \tref{tab:numbers}. Concerning the magnitude of the ordering vector there is some disagreement between the LMTO and SKKR calculations for the FM reference state. This can easily happen due to the delicate interplay of frustrated interactions we have demonstrated.

The $J\left(\mathbf q\right)$ surfaces for these two spin models plotted in the entire 2d BZ in \fref{fig:jq_surfaces} reflect the frustrated nature of the exchange interactions. In both cases there is a nearly degenerate line along $\Gamma - {\rm K}$, and this degeneracy is especially pronounced in case of the LMTO spin model. While the overall structure of the two surfaces is very similar, the numerical maximum is at an inner point of the $\Gamma - {\rm K}$ line for the SKKR couplings, whereas it is pushed out into the ${\rm K}$ point for the LMTO couplings. This is the reason for the difference in the mean-field ordering vectors seen in \tref{tab:numbers}.

Furthermore, considering the respective maximum $J\left(\mathbf q\right)$ values, the mean-field Néel temperature is unreasonably high compared to experiments, even with accounting for the expected overestimation.
Due to the much faster decay of the exchange interactions derived in the DLM state, the mean-field value of the ordering vector is nearly the same in the two \emph{ab initio} methods. Remarkably, the mean-field estimates for the Néel temperature from the DLM spin models are less than half of those from the FM spin models.

\begin{table}[htb]
  \caption{\label{tab:numbers}Estimates for the magnetic ordering. MF refers to the mean-field theory, while SD and MC refer to Landau--Lifshitz--Gilbert spin dynamics and Monte Carlo simulations, respectively.
Calculated Néel temperatures are in K, ordering vectors are expressed in units of $2\pi/(a/2)$ along $\Gamma-{\rm K}$ (such that $\rm K$ is at 1/3). The Monte Carlo critical temperature is estimated from an inverse power-law fit to the peak of the specific heat from the left and right side, and the uncertainties are estimated from the disagreement between the temperatures obtained from the pair of fits for a given method. The experimental ordering vector occurs at $q_0^{\rm exp}=0.285$; the experimental ordering temperature is 88 K \cite{funahashi1977}.
}
  \begin{indented}
    \item[]
      \begin{tabular}{@{}llllll}
        \br
        reference            & method      & $q_0^{\rm MF}$ & $T_{\rm N}^{\rm MF}$ & $q_0^{\rm SD}$ & $T_{\rm N}^{\rm MC}$ \\\hline\hline
        \mr
        \multirow{2}{*}{FM}  & LMTO        & 0.333          & 475                  & 0.330          & 187 ($\pm 3$) \\
                             & SKKR        & 0.208          & 439                  & 0.210          & 221 ($\pm 2$) \\
        \hline
        \multirow{2}{*}{DLM} & LMTO        & 0.210          & 204                  & 0.210          & 93 ($\pm 1$) \\
                             & SKKR        & 0.213          & 197                  & 0.213          & 97 ($\pm 1$) \\
        \hline
        \br
      \end{tabular}
  \end{indented}
\end{table}

In order to verify the mean-field estimates and to obtain a realistic temperature scale we performed Landau--Lifshitz--Gilbert spin dynamics simulations at zero temperature to find the ground-state
spin configuration, and Monte Carlo simulations at finite temperature to locate the Néel transition. As shown in \tref{tab:numbers}, the spatial modulation of the ground state estimate from spin dynamics
is in perfect agreement with the mean-field guess. With exception of the FM LMTO spin model, the magnitude of this wave vector of about $0.21 \times (2\pi)/(a/2)$ is significantly
smaller than seen in experiments and Ref.~\cite{brasse2013}.  We attribute this inaccuracy to the significant frustration in this system; the combination of exchange interactions causing the spin-spiral ground state is quite delicate and depends sensitively on details of the theory, e.g.\ the exchange-correlation model.  We note also that our calculations for the ordering vector are based on a bilinear spin model, missing higher-order multi-site exchange interactions, while methods based on total energy calculations, i.e.\ spin-spiral calculations~\cite{brasse2013}, do not involve this restriction.

\begin{figure}
  \begin{centering}
    \includegraphics[width=\linewidth]{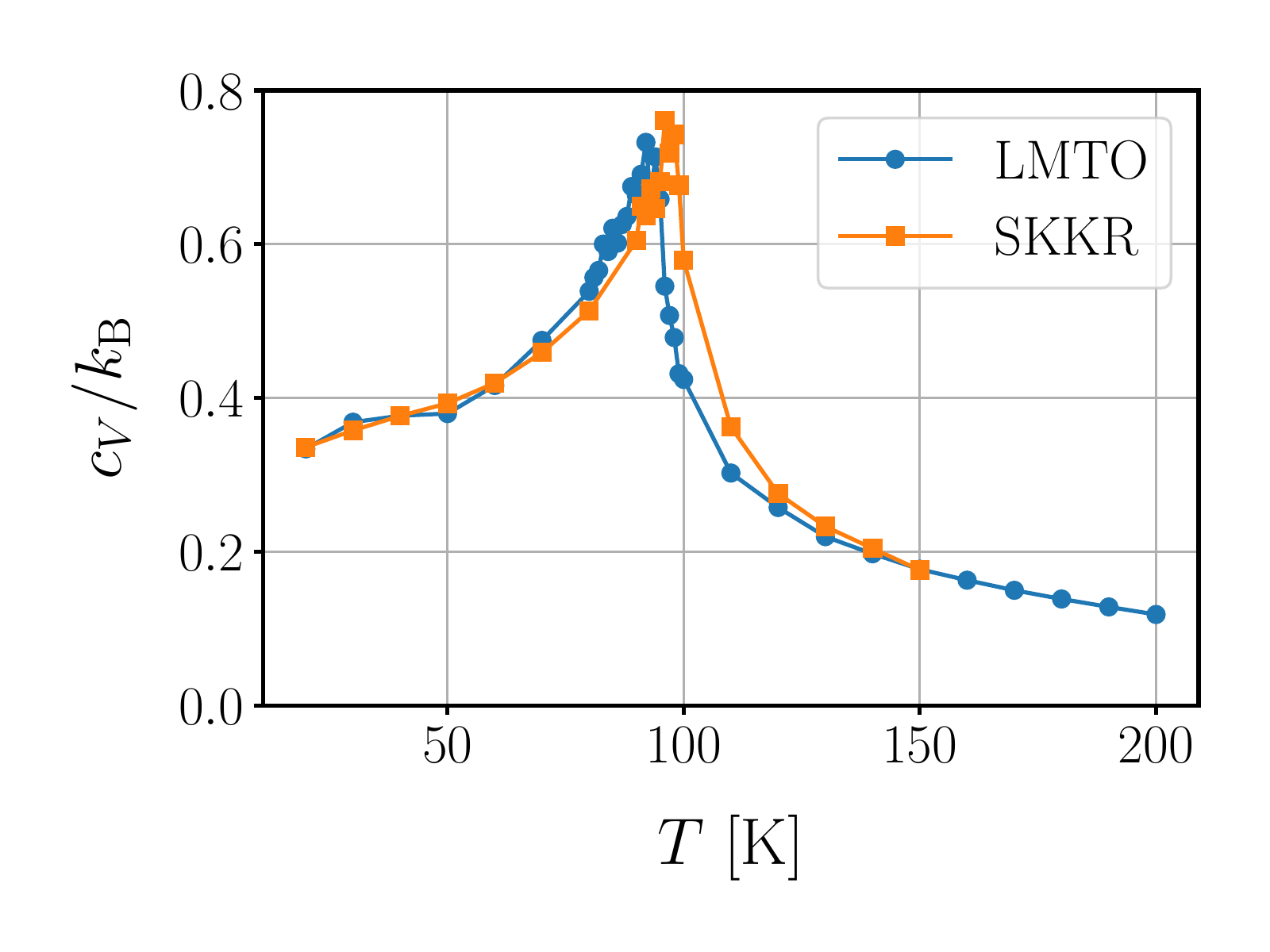}
  \end{centering}
  \caption{\label{fig:monte_carlo}Specific heat in Boltzmann's constant units as a function of temperature obtained from Monte Carlo simulations in a system of $32 \times 32 \times 32$ spins.}
\end{figure}

The temperature dependence of the specific heat from the Monte Carlo simulations of the DLM derived spin models is shown in \fref{fig:monte_carlo}. The sharp maximum of $c_V$ clearly indicates the position of the transition temperature. The about 97 K Néel temperature we obtained is in good agreement with experimental findings. This agreement is not surprising since the DLM theory describes by construction the magnetic interactions at the critical temperature, whereas the exchange interactions derived by the torque method at the FM state are essentially to describe the low-temperature spin-wave spectra. Indeed, the FM SKKR spin model resulted in a much too high Néel temperature of $221 \, {\rm K}$.

We note an interesting feature of our four spin models, namely that there is a striking 2.5 factor difference between the mean field estimate and the Monte Carlo result for the FM LMTO couplings, whereas for the other three spin models there is an almost exact factor of 2 instead. This dichotomy can probably be explained by the difference in ordered magnetic states: the triangular Néel structure (for the FM LMTO spin model) and a spin spiral with about 0.21 $2\pi/(a/2)$ wave number (for the other three spin models) may have different excitations, and thus significantly different fluctuations near the transition temperature.

\section{Conclusions}
In this work we presented Heisenberg exchange parameters describing the magnetism of CrB$_2$, which is known to have an incommensurate spin-spiral ground state.  By modelling the system in the paramagnetic phase using the disordered local moment theory  as implemented in the SKKR and LMTO Green's function methods of density functional theory, we have reproduced the essential magnetic properties of the system without making any assumptions as to the nature of the magnetic ordering at lower temperature.
In particular, the Néel temperature obtained with this theory is very close to that which is experimentally observed.  The ordering direction is correctly reproduced by the spin model,
but the wavelength of the ordering is underestimated.
We demonstrate also that using the ferromagnetic state as reference results in significantly different exchange interactions which are substantially poorer than those derived in the paramagnetic phase.  These conclusions are supported by rigorous testing with respect to the number of interacting shells of magnetic atoms.

Previous full-potential methods using the spin-spiral approach have found minimum energy configurations in better agreement with the experimental ordering vector~\cite{brasse2013}. This difference might be attributed either to the fundamental difference between our spin model based on mapping the band energy in the spirit of the force theorem \cite{LKAG1987,turek2006,szunyogh2011}  and the spin-spiral method based on total energy calculations or to the imprecision caused by the simplified representation of the Kohn--Sham potential in the ASA Green's function methods used here.  In return for the modest loss of precision inherent in the ASA, the methods employed here allow a detailed mapping of the exchange interactions over many neighbour shells and highlight the complex frustration driving incommensurate ordering in this material. The resulting spin model correctly describes the thermodynamics of the spin system, despite the relatively small moment and small ordering temperature. It is possible that an improvement could be achieved by performing the RTM calculations with a non-collinear reference state that closely resembles the ground state of the system, however there is a strong debate still in the literature on the calculation of spin models from non-collinear reference states \cite{cardias2020,dias2021,cardias2022,dias2022}.

\ack
AD, BN and LS acknowledge support by
National Research, Development and Innovation (NRDI) Office of Hungary under grant No.\ K131938,  PD134579 and TKP2021-NVA-02.
JJ acknowledges support under the CCP9 project ``Computational
Electronic Structure of Condensed Matter'' (part of the Computational Science Centre for Research Communities (CoSeC)).
Computing resources provided by STFC Scientific Computing Department's SCARF cluster,
and Governmental Information Technology Development Agency's (KIFÜ) cluster in Debrecen, Hungary.
Our spin simulation and data visualization tooling makes heavy use of the NumPy, Matplotlib, Pythran and PyVista open-source Python packages \cite{harris2020,*hunter2007,*guelton2015,*sullivan2019}.

\section*{References}
\bibliography{crb2}

\begin{thebibliography}{10}

\bibitem{laszloffy2019}
A.~L\'aszl\'offy, L.~R\'ozsa, K.~Palot\'as, L.~Udvardi, and L.~Szunyogh.
\newblock Magnetic structure of monatomic {Fe} chains on {Re(0001)}: Emergence
  of chiral multispin interactions.
\newblock {\em Phys. Rev. B}, 99:184430, May 2019.

\bibitem{laszloffy2021}
András Lászlóffy, Krisztián Palotás, Levente Rózsa, and László
  Szunyogh.
\newblock Electronic and magnetic properties of building blocks of {Mn} and
  {Fe} atomic chains on {Nb(110)}.
\newblock {\em Nanomaterials}, 11(8), 2021.

\bibitem{rozsa2015}
Levente R\'ozsa, L\'aszl\'o Udvardi, L\'aszl\'o Szunyogh, and Istv\'an~A.
  Szab\'o.
\newblock Magnetic phase diagram of an {Fe} monolayer on {W(110)} and {Ta(110)}
  surfaces based on ab initio calculations.
\newblock {\em Phys. Rev. B}, 91:144424, Apr 2015.

\bibitem{dupe2014}
Bertrand Dup{\'e}, Markus Hoffmann, Charles Paillard, and Stefan Heinze.
\newblock Tailoring magnetic skyrmions in ultra-thin transition metal films.
\newblock {\em Nature Communications}, 5(1):4030, Jun 2014.

\bibitem{dupe2016}
B.~Dup{\'e}, G.~Bihlmayer, M.~B{\"o}ttcher, S.~Bl{\"u}gel, and S.~Heinze.
\newblock Engineering skyrmions in transition-metal multilayers for
  spintronics.
\newblock {\em Nature Communications}, 7(1):11779, Jun 2016.

\bibitem{rozsa2016}
Levente R\'ozsa, Andr\'as De\'ak, Eszter Simon, Rocio Yanes, L\'aszl\'o
  Udvardi, L\'aszl\'o Szunyogh, and Ulrich Nowak.
\newblock Skyrmions with attractive interactions in an ultrathin magnetic film.
\newblock {\em Phys. Rev. Lett.}, 117:157205, Oct 2016.

\bibitem{park2018}
Pyeongjae Park, Joosung Oh, Kl{\'a}ra Uhl{\'i}{\v{r}}ov{\'a}, Jerome Jackson,
  Andr{\'a}s De{\'a}k, L{\'a}szl{\'o} Szunyogh, Ki~Hoon Lee, Hwanbeom Cho,
  Ha-Leem Kim, Helen~C. Walker, Devashibhai Adroja, Vladim{\'i}r Sechovsk{\'y},
  and Je-Geun Park.
\newblock Magnetic excitations in non-collinear antiferromagnetic weyl
  semimetal {Mn$_3$Sn}.
\newblock {\em npj Quantum Materials}, 3(1):63, Dec 2018.

\bibitem{portnoi1969}
K.~I. Portnoi, V.~M. Romashov, and I.~V. Romanovich.
\newblock Diagram of state of the chromium-boron system.
\newblock {\em Soviet Powder Metallurgy and Metal Ceramics}, 8(4):298--302,
  April 1969.

\bibitem{funahashi1977}
S.~Funahashi, Y.~Hamaguchi, T.~Tanaka, and E.~Bannai.
\newblock Helical magnetic structure in {CrB}$_{\textrm{2}}$.
\newblock {\em Solid State Communications}, 23(11):859--862, September 1977.

\bibitem{park2020}
Pyeongjae Park, Kisoo Park, Taehun Kim, Yusuke Kousaka, Ki~Hoon Lee, T.~G.
  Perring, Jaehong Jeong, Uwe Stuhr, Jun Akimitsu, Michel Kenzelmann, and
  Je-Geun Park.
\newblock Momentum-{Dependent} {Magnon} {Lifetime} in the {Metallic}
  {Noncollinear} {Triangular} {Antiferromagnet} {CrB}$_2$.
\newblock {\em Physical Review Letters}, 125(2):027202, July 2020.

\bibitem{kaya2009}
E.~Kaya, Y.~Kousaka, K.~Kakurai, M.~Takeda, and J.~Akimitsu.
\newblock Spherical neutron polarimetry studies on the magnetic structure of
  single crystal {Cr$_{1-x}$Mo$_x$B$_2$ ($x=0$, $0.15$)}.
\newblock {\em Physica B: Condensed Matter}, 404(17):2524--2526, September
  2009.

\bibitem{bauer2014}
A.~Bauer, A.~Regnat, C.~G.~F. Blum, S.~Gottlieb-Schönmeyer, B.~Pedersen,
  M.~Meven, S.~Wurmehl, J.~Kuneš, and C.~Pfleiderer.
\newblock Low-temperature properties of single-crystal {CrB}$_{\textrm{2}}$.
\newblock {\em Physical Review B}, 90(6):064414, August 2014.

\bibitem{liu1975}
S.~H. Liu, L.~Kopp, W.~B. England, and H.~W. Myron.
\newblock Energy bands, electronic properties, and magnetic ordering of
  {Cr}{B}$_{\textrm{2}}$.
\newblock {\em Physical Review B}, 11(9):3463--3468, May 1975.

\bibitem{vajeeston2001}
P.~Vajeeston, P.~Ravindran, C.~Ravi, and R.~Asokamani1.
\newblock Electronic structure, bonding, and ground-state properties of
  {AlB}$_{\textrm{2}}$-type transition-metal diborides.
\newblock {\em Physical Review B}, 63(4):045115, January 2001.

\bibitem{grechnev2009}
G.E. Grechnev, A.V. Fedorchenko, A.V. Logosha, A.S. Panfilov, I.V. Svechkarev,
  V.B. Filippov, A.B. Lyashchenko, and A.V. Evdokimova.
\newblock Electronic structure and magnetic properties of transition metal
  diborides.
\newblock {\em Journal of Alloys and Compounds}, 481(1-2):75--80, July 2009.

\bibitem{brasse2013}
M.~Brasse, L.~Chioncel, J.~Kuneš, A.~Bauer, A.~Regnat, C.~G.~F. Blum,
  S.~Wurmehl, C.~Pfleiderer, M.~A. Wilde, and D.~Grundler.
\newblock de {Haas}–van {Alphen} effect and {Fermi} surface properties of
  single-crystal {CrB}$_{\textrm{2}}$.
\newblock {\em Physical Review B}, 88(15):155138, October 2013.

\bibitem{gyorffy1985}
B~L Gyorffy, A~J Pindor, J~Staunton, G~M Stocks, and H~Winter.
\newblock A first-principles theory of ferromagnetic phase transitions in
  metals.
\newblock {\em Journal of Physics F: Metal Physics}, 15(6):1337--1386, Jun
  1985.

\bibitem{kkrbook}
J.~Zabloudil, R.~Hammerling, L.~Szunyogh, and P.~Weinberger.
\newblock {\em Electron Scattering in Solid Matter}.
\newblock Springer, Berlin, Heidelberg, Heidelberg, 2005.

\bibitem{perdew1996}
John~P. Perdew, Kieron Burke, and Matthias Ernzerhof.
\newblock Generalized gradient approximation made simple.
\newblock {\em Physical Review Letters}, 77:3865--3868, Oct 1996.

\bibitem{udvardi2003}
L.~Udvardi, L.~Szunyogh, K.~Palot\'as, and P.~Weinberger.
\newblock First-principles relativistic study of spin waves in thin magnetic
  films.
\newblock {\em Physical Review B}, 68:104436, Sep 2003.

\bibitem{staunton2006}
J.~B. Staunton, L.~Szunyogh, A.~Buruzs, B.~L. Gyorffy, S.~Ostanin, and
  L.~Udvardi.
\newblock Temperature dependence of magnetic anisotropy: An ab initio approach.
\newblock {\em Phys. Rev. B}, 74:144411, Oct 2006.

\bibitem{szunyogh2011}
L.~Szunyogh, L.~Udvardi, J.~Jackson, U.~Nowak, and R.~Chantrell.
\newblock Atomistic spin model based on a spin-cluster expansion technique:
  Application to the {IrMn${}_{3}$/Co} interface.
\newblock {\em Phys. Rev. B}, 83:024401, Jan 2011.

\bibitem{pashov2020}
D.~Pashov, S.~Acharya, W.R.L. Lambrecht, J.~Jackson, K.~D. Belashchenko,
  A.~Chantis, F.~Jamet, and M.~van Schilfgaarde.
\newblock Questaal: {A} package of electronic structure methods based on the
  linear muffin-tin orbital technique.
\newblock {\em Computer Physics Communications}, 249:107065, April 2020.

\bibitem{nowak2007}
Ulrich Nowak.
\newblock {\em Handbook of Magnetism and Advanced Magnetic Materials, Volume 2
  Micromagnetism}, chapter Classical Spin Models.
\newblock Wiley, New York, 2007.

\bibitem{LKAG1987}
A.I. Liechtenstein, M.I. Katsnelson, V.P. Antropov, and V.A. Gubanov.
\newblock Local spin density functional approach to the theory of exchange
  interactions in ferromagnetic metals and alloys.
\newblock {\em Journal of Magnetism and Magnetic Materials}, 67(1):65--74,
  1987.

\bibitem{turek2006}
I.~Turek, J.~Kudrnovský, V.~Drchal, and P.~Bruno.
\newblock Exchange interactions, spin waves, and transition temperatures in
  itinerant magnets.
\newblock {\em Philosophical Magazine}, 86(12):1713--1752, April 2006.

\bibitem{deak2011}
A.~De\'ak, L.~Szunyogh, and B.~Ujfalussy.
\newblock Thickness-dependent magnetic structure of ultrathin {Fe}/{Ir}(001)
  films: From spin-spiral states toward ferromagnetic order.
\newblock {\em Phys. Rev. B}, 84:224413, Dec 2011.

\bibitem{cardias2020}
R.~Cardias, A.~Szilva, M.~M. Bezerra-Neto, M.~S. Ribeiro, A.~Bergman, Y.~O.
  Kvashnin, J.~Fransson, A.~B. Klautau, O.~Eriksson, and L.~Nordstr{\"o}m.
\newblock First-principles {Dzyaloshinskii--Moriya} interaction in a
  non-collinear framework.
\newblock {\em Scientific Reports}, 10(1):20339, Nov 2020.

\bibitem{dias2021}
Manuel dos Santos~Dias, Sascha Brinker, Andr\'as L\'aszl\'offy, Bendeg\'uz
  Ny\'ari, Stefan Bl\"ugel, L\'aszl\'o Szunyogh, and Samir Lounis.
\newblock Proper and improper chiral magnetic interactions.
\newblock {\em Phys. Rev. B}, 103:L140408, Apr 2021.

\bibitem{cardias2022}
Ramon Cardias, Attila Szilva, Anders Bergman, Yaroslav Kvashnin, Jonas
  Fransson, Simon Streib, Anna Delin, Mikhail~I. Katsnelson, Danny Thonig,
  Angela~Burlamaqui Klautau, Olle Eriksson, and Lars Nordstr\"om.
\newblock {Comment on ``Proper and improper chiral magnetic interactions''}.
\newblock {\em Phys. Rev. B}, 105:026401, Jan 2022.

\bibitem{dias2022}
Manuel dos Santos~Dias, Sascha Brinker, Andr\'as L\'aszl\'offy, Bendeg\'uz
  Ny\'ari, Stefan Bl\"ugel, L\'aszl\'o Szunyogh, and Samir Lounis.
\newblock {Reply to ``Comment on `Proper and improper chiral magnetic
  interactions' ''}.
\newblock {\em Phys. Rev. B}, 105:026402, Jan 2022.

\bibitem{harris2020}
Charles~R. Harris, K.~Jarrod Millman, St\'efan~J. van~der Walt, Ralf Gommers,
  Pauli Virtanen, David Cournapeau, Eric Wieser, Julian Taylor, Sebastian Berg,
  Nathaniel~J. Smith, Robert Kern, Matti Picus, Stephan Hoyer, Marten~H. van
  Kerkwijk, Matthew Brett, Allan Haldane, Jaime~Fern\'andez del R\'io, Mark
  Wiebe, Pearu Peterson, Pierre G\'erard-Marchant, Kevin Sheppard, Tyler Reddy,
  Warren Weckesser, Hameer Abbasi, Christoph Gohlke, and Travis~E. Oliphant.
\newblock Array programming with {NumPy}.
\newblock {\em Nature}, 585(7825):357, Sep 2020.

\bibitem{hunter2007}
J.~D. Hunter.
\newblock Matplotlib: A {2D} graphics environment.
\newblock {\em Computing in Science \& Engineering}, 9:90, 2007.

\bibitem{guelton2015}
Serge Guelton, Pierrick Brunet, Mehdi Amini, Adrien Merlini, Xavier Corbillon,
  and Alan Raynaud.
\newblock Pythran: Enabling static optimization of scientific python programs.
\newblock {\em Computational Science \& Discovery}, 8(1):014001, 2015.

\bibitem{sullivan2019}
Bane Sullivan and Alexander Kaszynski.
\newblock {PyVista: 3D plotting and mesh analysis through a streamlined
  interface for the Visualization Toolkit (VTK)}.
\newblock {\em Journal of Open Source Software}, 4(37):1450, May 2019.

\end{thebibliography}

\end{document}